\documentclass[12pt]{article}

\usepackage{amssymb}
\usepackage{amsxtra}
\usepackage{amsmath}
\usepackage{amsfonts}
\usepackage{CJK}
\usepackage[titletoc]{appendix}
\usepackage{graphicx}
\usepackage{color}
\numberwithin{equation}{section}

\newtheorem{thm}{Theorem}[section]

\newtheorem{prop}[thm]{Proposition}
\newtheorem{assumption}[thm]{Assumption}

\newtheorem{lem}[thm]{Lemma}

\newtheorem{rem}[thm]{Remark}

\newcommand{\eqa}{\begin{eqnarray}}
\newcommand{\eeqa}{\end{eqnarray}}
\newcommand{\beq}{\begin{equation}}
\newcommand{\eeq}{\end{equation}}
\newcommand{\nn}{\nonumber}

\allowdisplaybreaks

\begin{document}
\title{ { The coupled Fokas-Lenells equations by a Riemann-Hilbert approach}}

\author{{\footnotesize { Bei-bei Hu$^{a,b}$ , Tie-cheng Xia$^{a,}$\thanks{Corresponding author. E-mails: xiatc@shu.edu.cn(Tiecheng Xia); hubsquare@chzu.edu.cn(Beibei Hu)}}}\\
{\footnotesize { {\it$^{a}$Department of Mathematics, Shanghai University, Shanghai 200444, China }}}\\
{\footnotesize { \it $^{b}$School of Mathematicas and Finance, Chuzhou University, Anhui, 239000, China}}
}
\date{\small }\maketitle

\textbf{Abstract}: In this paper, we use the unified transform method to consider the initial-boundary value problem for the coupled Fokas-Lenells equations on the half-line, assuming that the solution $\{q(x,t),r(x,t)\}$ of the coupled Fokas-Lenells equations exists, we show that $\{q_x(x,t),r_x(x,t)\}$ can be expressed in terms of the unique solution of a matrix Riemann-Hilbert problem formulated in the plane of the complex spectral parameter $\lambda$. Thus, the solution $\{q(x,t),r(x,t)\}$ can be obtained by integration with respect to $x$.\\
\\
\textbf{Keywords} {Riemann-Hilbert problem; Coupled Fokas-Lenells equations; Unified transform method; Initial-boundary value problem}\\
\textbf{ PACS numbers} {02.30.Ik, 02.30.Jr, 03.65.Nk}\\
\textbf{ MSC(2010)}  {35G31, 35Q15, 35Q51}

\section{Introduction}
In 1995, Fokas \cite{Fokas1995} used bi-Hamiltonian method presented an integrable generalization of the nonlinear Schr\"{o}dinger (NLS in brief) equation as follows
\eqa iu_t-vu_{xt}+\gamma u_{xx}+\epsilon |u|^2(u+ivu_x)=0,\quad \epsilon=\pm1, \label{slisp}\eeqa
where $u=u(x,t)$ is a complex valued function, $\gamma$ and $v$ are nonzero real parameters. The Eq.(1.1) is a completely integrable equation
that is called Fokas-Lenells (FL in brief) equation, and when $v=0$ which it can be reduces to the NLS equation.
Lenells show that the FL equation arises as a model for nonlinear wave propagation in monomode optical fibers in\cite{Lenells2009}.
And he find that the FL equation is related to the NLS equation in the same way as the Camassa-Holm equation is related to the KdV equation
from the perspective of bi-Hamiltonian. Furthermore, Lenells and Fokas\cite{lenells2009} used the bi-Hamiltonian method to obtain the first
few conservation laws of Eq.(1.1) and derived its Lax pair, and used the Lax pair to solve the initial value problem and analyse solitons.
The FL equation was studied in a number of papers\cite{l2010,Vek2011,Lu2013,Mat2012} and their references, the breathers and
rogue waves of the FL equation given by the Darboux transformation (DT in brief) method \cite{he2012}. In addition, the long-time asymptotic behavior
of the solution of the FL equation by the Defit-Zhou method\cite{XU2015,Chen2017}.

Since the interaction of waves of different frequencies gives rise to two-component NLS equation, their multi-component generalizations have attracted
much attention. The most famous example might be Manakov's equation, which is characterized by equal nonlinear interaction between two components.
Similarly, there are another recent integrable generalization of Manakov's equation for describing the effects of polarization or anisotropy, which is called the coupled Fokas-Lenells (CFL in brief) system \cite{Zhang2015,Zhang2017}, is given by
\beq
\left(\begin{array}{c}
p_1\\
p_2\\
r_1\\
r_2\end{array} \right)_t=i\left(\begin{array}{c}
\gamma u_{1,xx}-2p_1u_1v_1-p_1u_2v_2-p_2u_1v_2\\
\gamma u_{2,xx}-2p_2u_2v_2-p_2u_1v_1-p_1u_2v_1\\
-\gamma v_{1,xx}+2r_1u_1v_1+r_1u_2v_2+r_2u_1v_2\\
-\gamma v_{2,xx}+2r_2u_2v_2+r_2u_1v_1+r_1u_2v_1\end{array} \right),\label{slisp}\eeq
$$ p_k=u_k+ivu_{k,x},\quad r_k=v_k+ivv_{k,x},\; k=1,2,$$
where $u_k(x,t),v_k(x,t)$ is a complex valued function, $\gamma$ and $v$ are nonzero real parameters.
In fact, taking $v_k=\epsilon \overline u_k,\epsilon=\pm1$, the Eq.(1.2) can be written in the following form:
\beq \left\{\begin{array}{l}
iu_{1,t}-vu_{1,xt}+\gamma u_{1,xx}+\epsilon(2|u_1|^2+|u_2|^2)(u_1+ivu_{1,x})+\epsilon u_1\overline u_2(u_2+ivu_{2,x})=0.\\
iu_{2,t}-vu_{2,xt}+\gamma u_{2,xx}+\epsilon(2|u_2|^2+|u_1|^2)(u_2+ivu_{2,x})+\epsilon u_2\overline u_1(u_1+ivu_{1,x})=0,
\end{array}\right. \label{slisp}\eeq
where asterisk denotes the complex conjugation.

Moreover, the Eq.(1.3) can also be written by a simple change of variables combined with a gauge transformation\cite{Lenells2009}
($u_1=e^{ix}q,u_2=e^{ix}r$) and the condition $\gamma=2,v=1,\epsilon=-1$ in the following CFL equations:
\eqa \left\{\begin{array}{l}
iq_{xt}-2iq_{xx}+4q_{x}-(2|q|^2+|r|^2)q_{x}-q\overline rr_{x}+2iq=0,\\
ir_{xt}-2ir_{xx}+4r_{x}-(2|r|^2+|q|^2)r_{x}-r\overline qq_{x}+2ir=0.
\end{array}\right. \label{slisp}\eeqa
Most recently, the CFL equation has been studied by several authors. Such as the infinite conservation laws of the CFL equations has been studied in \cite{Ling2018}, and the higher-order soliton, breather, and rogue wave solutions of the CFL equations are derived via the n-fold Darboux transformation in \cite{Zhang2017}.

In 1997, Fokas structured a new unified approach for the analysis of initial-boundary value (IBV in brief) problems for linear and nonlinear integrable partial differential equations (PDEs in brief) \cite{Fokas1997,Fokas2002,Fokas2008}, we call that unified transform method. This method provides an important generalization of the inverse scattering transform (IST in brief) formalism from initial value to IBV problems, and over the last 20 years, this method has been used to analyse boundary value problems for several of the most important integrable equations possessing $2\times2$ Lax pairs, such as the KdV, the NLS, the sine-Gordon \cite{Lenells2011,Fokas2005,Len2011} and others \cite{Deconinck2017,hu2017,Xiao2017,Xia2017,ZhangN2017}. Just like the IST on the line, the unified transform method yields an expression for the solution of an IBV problem in terms of the solution of a Riemann-Hilbert problem. In particular, an effective way analyzing the asymptotic behaviour of the solution is based on this Riemann-Hilbert problem and by employing the nonlinear version of the steepest descent method introduced by Deift and Zhou \cite{Deift1993}.

In 2012, Lenells first extended the Fokas unified transform method to the IBV problem for the $3\times3$ matrix Lax pair \cite{Lenells2012,Lenells2013}. After that, more and more researchers begin to pay attention to studying IBV problems for integrable evolution equations with higher order Lax pairs on the half-line or on the interval, the IBV problem for the many integrable equations with $3\times3$ or $4\times4$ Lax pairs are studied, such as, the Degasperis-Procesi equation \cite{Lenells2013,Monvel2013}, the Ostrovsky-Vakhnenko equation \cite{Monvel2014,Monvel2015}, the Sasa-Satsuma equation \cite{Xu2013}, the three wave equation \cite{Xu2014}, the coupled NLS equation \cite{Geng2015}, the vector modified KdV equation \cite{Liu2016}, the Novikov equation \cite{Monvel2016}, the integrable spin-1 Gross-Pitaevskii equations with a $4\times4$ Lax pair \cite{Yan2017} and others \cite{Hu2017,Hu2018,Tian2017,Xu2016,Zhu2017}. We have a good time to study PDEs with IBV problem, and has also done some work about integrable equations with $2\times2$ or $3\times3$ Lax pairs on the half-line \cite{Zhang2017,Hu2018,Hu2017}.

Likewise, here our aim is implement the unified transform method to analyze the IBV problem for the CFL equations (1.4), and the initial boundary values datas lie in the Schwartz class defined by
\eqa\begin{array}{l}
$Initial values$: q_0(x)=u(x,t=0),\; r_0(x)=v(x,t=0),\;0<x<\infty;\\
$Dirichlet boundary values$: g_0(t)=q(x=0,t),\; h_0(t)=r(x=0,t),\;0<t<T;\\
$Neumann boundary values$: g_1(t)=q_x(x=0,t),\; h_1(t)=r_x(x=0,t),\;0<t<T.
\end{array}\label{slisp}\eeqa
In this work, we use the unified transform method to deal with this problem on the half-line $\Omega=\{0<x<\infty,0<t<T\}$. We assume that the solution $\{q(x,t),r(x,t)\}$ of CFL equations exists. Through this method, we show that $\{q_x(x,t),r_x(x,t)\}$ can be expressed in terms of the unique solution of a matrix Riemann-Hilbert problem formulated in the plane of the complex spectral parameter $\lambda$. Thus, the solution $\{q(x,t),r(x,t)\}$ CFL equations can be obtained by integration with respect to $x$.

This paper is organized as follows. In section 2, we define two sets of eigenfunctions $\mu_j(j=1,2,3)$ and $M_n(n=1,2,3,4)$ of Lax pair for spectral analysis. In section 3, we show that $q_x(x, t),r_x(x, t)$ can be expressed in terms of the unique solution of a matrix Riemann-Hilbert problem, and the solution $\{q(x,t),r(x,t)\}$ CFL equations can be obtained by integration with respect to $x$. The last section is devoted to conclusions and discussions.


\section{ The spectral analysis}

The coupled Fokas-Lenells equations (1.4) admits the Lax pair formulation \cite{Zhang2017}
 \eqa \left\{\begin{array}{l}
 \psi_x=(i\lambda^2\sigma+\lambda Q)\psi,\\
\psi_t=(2i\lambda^2\sigma+2\lambda Q-2i\sigma +iV_0+iV_{-1}\lambda^{-1}+\frac{1}{2}i\lambda^{-2}\sigma)\psi,
\end{array}\right.\label{slisp}\eeqa
where
\eqa \begin{array}{l}
 \sigma=\left(\begin{array}{ccc}
-1&0&0\\
0&1&0\\
0&0&1\end{array} \right),
Q=\left(\begin{array}{ccc}
0&q_x&r_x\\
\overline q_x&0&0\\
\overline r_x&0&0\end{array} \right),\\
V_{0}=\left(\begin{array}{ccc}
-|q|^2-|r|^2 & 0 & 0\\
0 & |q|^2 & \bar qr\\
0 & q\bar r & |r|^2\end{array} \right),
V_{-1}=\left(\begin{array}{ccc}
0&q&r\\
-\overline q&0&0\\
-\overline r&0&0\end{array} \right),
\end{array}\eeqa
It is not difficult to find that Eq.(2.1) is equivalent to
\eqa \left\{\begin{array}{l}
\psi_x-i\lambda^2\sigma=U_1\psi,\\
\psi_t-2ik^2\sigma=U_2\psi,\\
\end{array}\right.\label{slisp}\eeqa
where \beq k=\lambda-\frac{1}{2\lambda},\,U_1=\lambda Q,\, U_2=2\lambda Q+iV_0+iV_{-1}\lambda^{-1}.\eeq

\subsection{ The closed one-form}

Assume that $q(x,t),r(x,t)$ are sufficiently smooth function in the half-line region of $\Omega=\{0<x<\infty,0<t<T\}$, and decays sufficiently when $x\rightarrow\infty.$ Extend the column vector $\psi$ to a $3\times3$ matrix and introducing a new eigenfunction $\mu(x,t,\lambda)$ by
\beq\psi=\mu e^{i\lambda^2\sigma x+2ik^2\sigma t},\label{slisp}\eeq
then the Lax pair equation Eq.(2.3) becomes
\beq \left\{\begin{array}{l}
\mu_x-i\lambda^2[\sigma,\mu]=V_1\mu,\\
\mu_t-2ik^2[\sigma,\mu]=V_2\mu,\\
\end{array}\right.\label{slisp}\eeq
which can be written in full derivative form
\beq d(e^{-i\lambda^2\hat\sigma x-2ik^2\hat\sigma t}\mu)=W(x,t,\lambda), \label{slisp}\eeq
where
\beq W(x,t,\lambda)=e^{-(i\lambda^2 x+2ik^2t)\hat\sigma}(V_1dx+V_2dt)\mu, \label{slisp}\eeq
where $\hat\sigma$ acts on a $3\times3$ matrix A by $e^{\hat\sigma A}=e^{\sigma}Ae^{-\sigma}$ and $3\times3$ matrix B by  $\hat\sigma B=[\sigma,B]$.

\subsection{ The eigenfunction}

There are three eigenfunctions $\mu_j(x,t,\lambda)(j=1,2,3)$ of Eq.(2.6) are defined by the following the Volterra integral equation
\eqa &&\mu_j(x,t,\lambda)=\mathbb{I}+\int_{\gamma_j}e^{(i\lambda^2 x+2ik^2t)\hat\sigma}W_j(x,t,\lambda)\nn\\&&
\qquad\qquad =\mathbb{I}+\int_{(x_j,t_j)}^{(x,t)}e^{(i\lambda^2 x+2ik^2t)\hat\sigma}W_j(x,t,\lambda),\quad j=1,2,3,\label{slisp}\eeqa
where $W_j$ is given by Eq.(2.8), it is only used $\mu_j$ in place of $\mu$, the contours $\gamma_j(j=1,2,3)$ are shown in figure 1.
and $(x_1,t_1)=(0,T)$, $(x_2,t_2)=(0,0)$, and $(x_3,t_3)=(\infty,t)$.

\begin{figure}
\centering
\includegraphics[width=4.4in,height=1.4in]{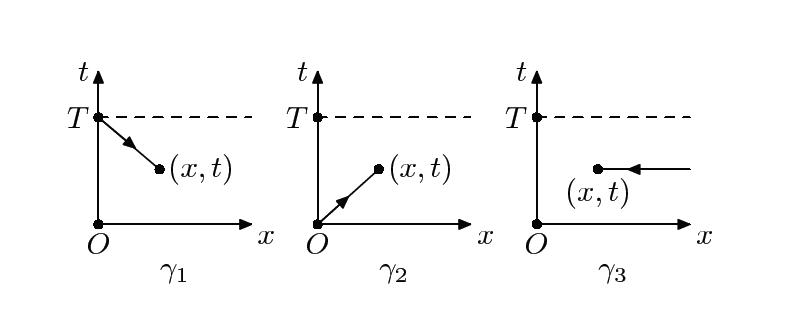}
\caption{The three contours $\gamma_1,\gamma_2,\gamma_3$ in the $(x,t)$-domaint}
\label{fig:graph}
\end{figure}
So we have that the following inequalities are hold true on the contours $\gamma_j(j=1,2,3)$
\eqa\begin{array}{l}
 \gamma_1=(x_1,t_1)\rightarrow(x,t):\, x-\xi\geq 0,\, t-\tau\leq 0,\\
 \gamma_2=(x_2,t_2)\rightarrow(x,t):\, x-\xi\geq 0,\, t-\tau\geq 0,\\
 \gamma_3=(x_3,t_3)\rightarrow(x,t):\, x-\xi\leq 0.
\end{array}\label{slisp}\eeqa

Define the following sets (see Figure 2) as
\eqa\begin{array}{l}
 D_1=\{\lambda\in C|\mathrm{arg} \lambda\in(0,\frac{\pi}{2})\cup(\pi,\frac{3\pi}{2}) $and$ |\lambda|>\frac{1}{\sqrt2}\},\\
 D_2=\{\lambda\in C|\mathrm{arg} \lambda\in(0,\frac{\pi}{2})\cup(\pi,\frac{3\pi}{2}) $and$ |\lambda|<\frac{1}{\sqrt2}\},\\
 D_3=\{\lambda\in C|\mathrm{arg} \lambda\in(\frac{\pi}{2},\pi)\cup(\frac{3\pi}{2},2\pi) $and$ |\lambda|<\frac{1}{\sqrt2}\},\\
 D_4=\{\lambda\in C|\mathrm{arg} \lambda\in(\frac{\pi}{2},\pi)\cup(\frac{3\pi}{2},2\pi) $and$ |\lambda|>\frac{1}{\sqrt2}\}.
\end{array}\nn\eeqa
From Eq.(2.4), we have
$$k=\lambda-\frac{1}{2\lambda}=(1-\frac{1}{2|\lambda|^2})\mathrm{Re}\lambda+i(1+\frac{1}{2|\lambda|^2})\mathrm{Im}\lambda.$$

As the first, second, and third columns of the matrix equation Eq.(2.9) contain the following exponential term
\eqa\begin{array}{l}
\mu_j^{(1)}:\quad e^{-2i\lambda^2(x-\xi)-4ik^2t(t-\tau)},\quad e^{-2i\lambda^2(x-\xi)-4ik^2t(t-\tau)},\\
\mu_j^{(2)}:\quad e^{2i\lambda^2(x-\xi)+4ik^2t(t-\tau)},\\
\mu_j^{(3)}:\quad e^{2i\lambda^2(x-\xi)+4ik^2t(t-\tau)}.\\
\end{array}\label{slisp}\eeqa

Thus, these inequalities imply that the function $\mu_j(x,t,\lambda)(j=1,2,3)$ is bounded and analytic in the following regions
\eqa\begin{array}{l}
 \mu_1 \quad $is bounded for$\quad \lambda\in(D_4,D_2,D_2),\\
 \mu_2\quad $is bounded for$\quad \lambda\in(D_3,D_1,D_1),\\
 \mu_3 \quad $is bounded for$\quad \lambda\in(D_1\cup D_2,D_3\cup D_4,D_3\cup D_4),
\end{array}\label{slisp}\eeqa
where $D_n(n=1,2,3,4)$ represents a subset of four open disjoint $\lambda-$ plane shown in figure 2.

\begin{figure}
  \centering
  \includegraphics[width=3.0in,height=2.8in]{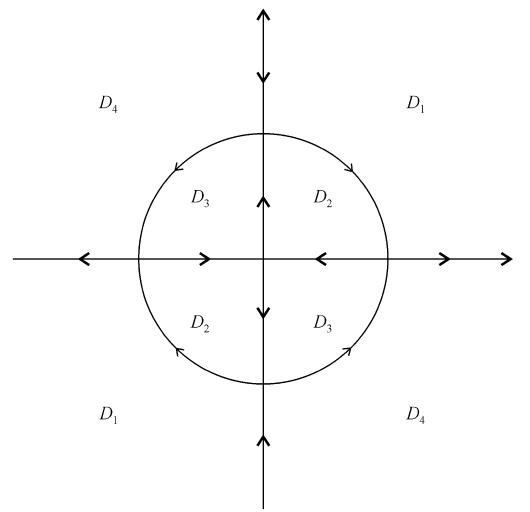}\\
  \caption{The sets $D_n,n=1,2,3,4$, which decompose the complex $\lambda-$plane}\label{fig:graph}
\end{figure}

And these sets $D_n(n=1,2,3,4)$ have the following properties
\eqa\begin{array}{l}
D_1=\{\lambda\in C |\mathrm{Re}l_1<\mathrm{Re}l_2=\mathrm{Re}l_3,\quad \mathrm{Re}z_1<\mathrm{Re}z_2=\mathrm{Re}z_3\},\\
D_2=\{\lambda\in C |\mathrm{Re}l_1<\mathrm{Re}l_2=\mathrm{Re}l_3,\quad \mathrm{Re}z_1>\mathrm{Re}z_2=\mathrm{Re}z_3\},\\
D_3=\{\lambda\in C |\mathrm{Re}l_1>\mathrm{Re}l_2=\mathrm{Re}l_3,\quad \mathrm{Re}z_1<\mathrm{Re}z_2=\mathrm{Re}z_3\},\\
D_4=\{\lambda\in C |\mathrm{Re}l_1>\mathrm{Re}l_2=\mathrm{Re}l_3,\quad \mathrm{Re}z_1>\mathrm{Re}z_2=\mathrm{Re}z_3\},
\end{array}\label{slisp}\eeqa
where $l_i(\lambda)$ and $z_i(\lambda)$ are the diagonal elements of the matrix $-i\lambda^2\sigma$ and $-2ik^2\sigma$.

In fact, $\mu_1(0,t,\lambda)$ has a larger bounded and analytic domain $(D_1\cup D_4,D_2\cup D_3,D_2\cup D_3)$ for $x=0$, and it is not difficult to see that $\mu_2(0,t,\lambda)$ has also a larger bounded and analytic domain $(D_2\cup D_3,D_1\cup D_4,D_1\cup D_4)$ for $x=0$.

\subsection{ The matrix value eigenfunction}

For each $n=1,2,3,4$, a solution $M_n(x,t,\lambda)$ of Eq.(2.6) is defined by the following integral equation
\beq (M_n(x,t,\lambda))_{ij}=\delta_{ij}+\int_{\gamma_{ij}^n}(e^{(i\lambda^2 x+2ik^2t)\hat\sigma}W_n(\xi,\tau,\lambda))_{ij},\quad i,j=1,2,3,\label{slisp}\eeq
where $W_n(x,t,\lambda)$ is given by Eq.(2.8), it is only used $M_n$ in place of $\mu$, and the contours $\gamma_{ij}^n (n=1,2,3,4;i,j=1,2,3)$ are defined as follows
 \eqa \gamma_{ij}^n=\left\{\begin{array}{l}\gamma_1\quad if\quad \mathrm{Re}l_i(\lambda)<\mathrm{Re}l_j(\lambda) \quad
 and \quad \mathrm{Re}z_i(\lambda)\geq \mathrm{Re}z_j(\lambda)\\
 \gamma_2\quad if\quad \mathrm{Re}l_i(\lambda)<\mathrm{Re}l_j(\lambda) \quad and \quad \mathrm{Re}z_i(\lambda)<\mathrm{Re}z_j(\lambda)\\
 \gamma_3\quad if \quad \mathrm{Re}l_i(\lambda)\geq \mathrm{Re}l_j(\lambda) \\
\end{array}\right. for \lambda\in D_n\quad \label{slisp}\eeqa
According to the definition of $\gamma^n$, we have
\eqa \begin{array}{l}
\gamma^1=\left(\begin{array}{ccc}
\gamma_3&\gamma_2&\gamma_2\\
\gamma_3&\gamma_3&\gamma_3\\
\gamma_3&\gamma_3&\gamma_3\end{array} \right),\,
\gamma^2=\left(\begin{array}{ccc}
\gamma_3&\gamma_1&\gamma_1\\
\gamma_3&\gamma_3&\gamma_3\\
\gamma_3&\gamma_3&\gamma_3\end{array} \right),\,\\
\gamma^3=\left(\begin{array}{ccc}
\gamma_3&\gamma_3&\gamma_3\\
\gamma_1&\gamma_3&\gamma_3\\
\gamma_1&\gamma_3&\gamma_3\end{array} \right),\,
\gamma^4=\left(\begin{array}{ccc}
\gamma_3&\gamma_3&\gamma_3\\
\gamma_2&\gamma_3&\gamma_3\\
\gamma_2&\gamma_3&\gamma_3\end{array} \right).
\end{array}\eeqa

Next, the following proposition guarantees that the previous definition of $M_n$ has properties that can be represented as a Rimann-Hilbert problem.

\begin{prop} For each $n=1,2,3,4$ and $\lambda\in D_n$, the function $M_n(x,t,\lambda)$ is defined well by Eq.(2.14). For any identified point $(x,t)$, $M_n$
is bounded and analytic as a function of $\lambda\in D_n$ away from a possible discrete set of singularities $\{\lambda_j\}$ at which the Fredholm determinant vanishes. Moreover, $M_n$ admits a bounded and continuous extension to $\bar D_n$ and
\beq M_n(x,t,\lambda)=\mathbb{I}+O(\frac{1}{\lambda}).\label{slisp}\eeq\end{prop}
Proof: The associated bounded and analytic properties have been established in Appendix B in \cite{Lenells2012}. Substituting the follow expansion
\beq M=M_0+\frac{M^{(1)}}{\lambda}+\frac{M^{(2)}}{\lambda^2}+\frac{M^{(3)}}{\lambda^3}+\frac{M^{(4)}}{\lambda^4}+\cdots,\quad\lambda\rightarrow\infty,\nn\eeq
into the Lax pair Eq.(2.6) and comparing the coefficients of $\lambda$ can obtain (2.7).

\subsection{ The jump matrix}
Define the spectral function as follows
\beq S_n(\lambda)=M_n(0,0,\lambda),\quad\lambda\in D_n,n=1,2,3,4.\label{slisp}\eeq
Let $M$ be a sectionally analytical continuous function in Riemann $\lambda-$ sphere which equals $M_n$ for
$\lambda\in D_n$. So, $M$ satisfies the following jump conditions
\beq M_n(\lambda)=M_mJ_{m,n},\quad \lambda\in \bar D_n\cap \bar D_m,\quad n,m=1,2,3,4;n\neq m,\label{slisp}\eeq
where
\beq J_{m,n}=e^{(-ik x+2ik^2t)\hat\Lambda}(S_m^{-1}S_n).\label{slisp}\eeq

\subsection{ The adjugated eigenfunction}

Likewise, we also need to consider the bounded and analytic properties of the minors of the matrices $\mu_j(x,t,\lambda)(j=1,2,3)$. We recall that the cofactor matrix $B^A$ of a $3\times3$ matrix $B$ is defined by
$$ B^A=\left(\begin{array}{ccc}
m_{11}(B)&-m_{12}(B)&m_{13}(B)\\
-m_{21}(B)&m_{22}(B)&-m_{23}(B)\\
m_{31}(B)&-m_{32}(B)&m_{33}(B) \end{array}
\right),$$
where $m_{ij}(B)$ denote the $(ij)$th minor of $B$.
From Eq.(2.6) we find that the adjugated eigenfunction $\mu^A$ satisfies the Lax pair
\eqa \left\{\begin{array}{l}
\mu_x^A+i\lambda^2[\sigma,\mu^A]=-V_1^T\mu^A,\\
\mu_t^A+2ik^2[\sigma,\mu^A]=-V_2^T\mu^A,
\end{array}\right.\label{slisp}\eeqa
where $V^T$ denotes the transform of a matrix $V$. So, the adjugated eigenfunctions $\mu_j(j=1,2,3)$ are solutions of the
integral equations
\beq \mu_j^A(x,t,\lambda)=I-\int_{\gamma_j}e^{(-i\lambda^2(x-\xi)-2ik^2(t-\tau))\hat\sigma}(V_1^Tdx+V_2^Tdt),\quad j=1,2,3,\label{slisp}\eeq
Thus, we can obtain the adjugated eigenfunction which satisfies the following analytic properties
\eqa\begin{array}{l}
 \mu_1^A\quad $is bounded for$\quad\lambda\in (D_2,D_4, D_4),\\
 \mu_2^A\quad $is bounded for$\quad\lambda\in (D_1,D_3,D_3),\\
 \mu_3^A\quad$is bounded for$\quad\lambda\in (D_3\cup D_4, D_1\cup D_2,D_1\cup D_2).
\end{array}\label{slisp}\eeqa

In fact, $\mu_1^A(0,t,\lambda)$ has a larger bounded and analytic domain which is $(D_2\cup D_3,D_1\cup D_4,D_1\cup D_4)$ for $x=0$, and $\mu_2^A(0,t,\lambda)$ also has a larger bounded analytic domain which is $(D_1\cup D_4,D_2\cup D_3,D_2\cup D_3)$ for $x=0$.

\subsection{ Symmetry}

By the following Lemma, we show that the eigenfunctions $\mu_j(x,t,\lambda)$ have an important symmetry.

\begin{lem} The eigenfunction $\psi(x,t,\lambda)$ of the Lax pair Eq.(2.1) admits the following symmetry
$$\psi^{-1}(x,t,\lambda)=A \overline{\psi(x,t,\bar\lambda)}^TA,$$
with
$$
A=\left(\begin{array}{ccc}
-1&0&0\\
0&\varepsilon&0\\
0&0&\varepsilon\end{array}
\right),\quad and \quad\varepsilon^2=1.$$
where the superscript T denotes a matrix transpose.\end{lem}

Proof: Analogous to the proof provided in \cite{Lenells2012}. We omit the proof.

\begin{rem} From Lemma 1, one can show that the eigenfunctions $\mu_j(x,t,\lambda)$ of Lax pair Eq.(2.6) have the same symmetry.\end{rem}

\subsection{ The jump matrix computation}

We define $3\times3$ matrix value function $s(\lambda)$ and $S(\lambda)$ as follows
\eqa\begin{array}{l}
\mu_3(x,t,\lambda)=\mu_2(x,t,\lambda)e^{(i\lambda^2 x+2ik^2t)\hat\sigma} s(\lambda),\\
\mu_1(x,t,\lambda)=\mu_2(x,t,\lambda)e^{(i\lambda^2 x+2ik^2t)\hat\sigma} S(\lambda),
\end{array}\label{slisp}\eeqa
as $\mu_2(0,0,\lambda)=\mathbb{I}$, we obtain
\beq s(\lambda)=\mu_3(0,0,\lambda),\quad S(\lambda)=\mu_1(0,0,\lambda).\label{slisp}\eeq
From the properties of $\mu_j$ and $\mu_j^A(j=1,2,3)$  we can obtain that $s(\lambda)$ and $S(\lambda)$ have the following bounded and analytic properties
\eqa\begin{array}{l} s(\lambda)\quad $is bounded for$\quad\lambda\in (D_1\cup D_2, D_3\cup D_4,D_3\cup D_4),\\
 S(\lambda)\quad $is bounded for$\quad\lambda\in (D_1\cup D_4, D_2\cup D_3,D_2\cup D_3),\\
 s^A(\lambda)\quad $is bounded for$\quad\lambda\in (D_3\cup D_4, D_1\cup D_2,D_1\cup D_2),\\
 S^A(\lambda)\quad$is bounded for$\quad\lambda\in (D_2\cup D_3, D_1\cup D_4,D_1\cup D_4),
\end{array}\label{slisp}\eeqa
moreover
\beq M_n(x,t,\lambda)=\mu_2(x,t,\lambda)e^{(i\lambda^2 x+2ik^2t)\hat\sigma} S_n(\lambda),\quad \lambda\in D_n.\label{slisp}\eeq
\begin{prop} $S_n$ can be expressed with $s(\lambda)$ and $S(\lambda)$  elements as follows
\eqa \begin{array}{l}
S_1=\left(\begin{array}{ccc}
s_{11}&0&0\\
s_{21}&\frac{m_{33}(s)}{s_{11}}&\frac{m_{32}(s)}{s_{11}}\\
s_{31}&\frac{m_{23}(s)}{s_{11}}&\frac{m_{22}(s)}{s_{11}}
\end{array} \right),\\
S_2=\left(\begin{array}{ccc}
s_{11}&\frac{m_{33}(s)M_{21}(S)-m_{23}(s)M_{31}(S)}{(s^TS^A)_{11}}&\frac{m_{32}(s)M_{21}(S)-m_{22}(s)M_{31}(S)}{(s^TS^A)_{11}}\\
s_{21}&\frac{m_{33}(s)M_{11}(S)-m_{13}(s)M_{31}(S)}{(s^TS^A)_{11}}&\frac{m_{32}(s)M_{11}(S)-m_{12}(s)M_{31}(S)}{(s^TS^A)_{11}}\\
s_{31}&\frac{m_{23}(s)M_{11}(S)-m_{13}(s)M_{21}(S)}{(s^TS^A)_{11}}&\frac{m_{22}(s)M_{11}(S)-m_{12}(s)M_{21}(S)}{(s^TS^A)_{11}}
\end{array} \right),\\
S_3=\left(\begin{array}{ccc}
\frac{S_{11}}{(S^Ts^A)_{11}}&s_{12}&s_{13}\\
\frac{S_{21}}{(S^Ts^A)_{11}}&s_{22}&s_{23}\\
\frac{S_{31}}{(S^Ts^A)_{11}}&s_{32}&s_{33}
\end{array} \right),
S_4=\left(\begin{array}{ccc}
\frac{1}{m_{11}(s)}&s_{12}&s_{13}\\
0&s_{22}&s_{23}\\
0&s_{32}&s_{33}
\end{array} \right),\end{array}\label{slisp}\eeqa
where $(S^Ts^A)_{11}$ and $(s^TS^A)_{11}$ are defined as follows
\eqa\begin{array}{l}
(S^Ts^A)_{11}=S_{11}m_{11}(s)-S_{21}m_{21}(s)+S_{31}m_{31}(s),\\
(s^TS^A)_{11}=s_{11}m_{11}(S)-s_{21}m_{21}(S)+s_{31}m_{31}(S).
\end{array}\nn\eeqa
\end{prop}
Proof: We set $\gamma_3^{X_0}$ is a contour when $(X_0,0)\rightarrow (x,t)$ in the $(x,t)$-plane, here $X_0$ is a constant and $X_0>0$, for $j=3$, we introduce $\mu_3(x,t,\lambda;X_0)$ as the solution of Eq.(2.9) with the contour $\gamma_3$ replaced by $\gamma_3^{X_0}$. Similarly, we define $M_n(x,t,\lambda;X_0)$ as the solution of Eq.(2.14) with $\gamma_3$ replaced by $\gamma_3^{X_0}$.
Then, by simple calculation, we can derive the expression of $S_n(\lambda,X_0)=M_n(0,0,\lambda;X_0)$ with $S(\lambda)$ and $s(\lambda;X_0)$ and the Eq.(2.28) will be obtain by taking the limit $X_0\rightarrow\infty$.

Firstly, we have the following relations:
\beq M_n(x,t,\lambda;X_0)=\mu_1(x,t,\lambda)e^{(i\lambda^2 x+2ik^2t)\hat\sigma} R_n(\lambda;X_0),\label{slisp}\eeq
\beq M_n(x,t,\lambda;X_0)=\mu_2(x,t,\lambda)e^{(i\lambda^2 x+2ik^2t)\hat\sigma} S_n(\lambda;X_0),\label{slisp}\eeq
\beq M_n(x,t,\lambda;X_0)=\mu_3(x,t,\lambda)e^{(i\lambda^2 x+2ik^2t)\hat\sigma} T_n(\lambda;X_0).\label{slisp}\eeq

Secondly, we get the definition of $R_n(\lambda;X_0)$ and $T_n(\lambda;X_0)$ as follows
\beq R_n(\lambda;X_0)=e^{-2ik^2T\hat\sigma} M_n(0,T,\lambda;X_0),\label{slisp}\eeq
\beq T_n(\lambda;X_0)=e^{-i\lambda^2 X_0\hat\sigma} M_n(X_0,0,\lambda;X_0),\label{slisp}\eeq
the Eq.(2.30) means that
\beq s(\lambda;X_0)=S_n(\lambda;X_0)T_n^{-1}(\lambda;X_0),\label{slisp}\eeq
\beq S(\lambda;X_0)=S_n(\lambda;X_0)R_n^{-1}(\lambda;X_0).\label{slisp}\eeq

These equations constitute the matrix decomposition problem of $\{s,S\}$ by use $\{R_n,S_n,T_n\}$. In fact,
by the definition of the integral equation Eq.(2.14) and $\{R_n,S_n,T_n\}$, we obtain
\eqa\left\{\begin{array}{l} (R_n(\lambda;X_0))_{ij}=0\quad if \quad \gamma_{ij}^n=\gamma_1,\\
(S_n(\lambda;X_0))_{ij}=0\quad if \quad \gamma_{ij}^n=\gamma_2,\\
(S_n(\lambda;X_0))_{ij}=\delta_{ij}\quad if \quad \gamma_{ij}^n=\gamma_3.
\end{array}\right.\label{slisp}\eeqa

Thus Eq.(2.31) are the eighteen scalar equations with eighteen unknowns. The exact solution of these system can be obtained by solving the algebraic system. In this way, we can get a similar $\{S_n(\lambda),s(\lambda)\}$ as in Eq.(2.28) which just that $\{S_n(\lambda),s(\lambda)\}$ replaces by $\{S_n(\lambda;X_0),s(\lambda;X_0)\}$ in Eq.(2.28).

Finally, taking $X_0\rightarrow\infty$ in this equation, we obtain the Eq.(2.28).

\subsection{ The residue conditions}

As $\mu_2$ is an entire function and by Eq.(2.27), it is not difficult to see that $M$ only produces singularities in $S_n$ where there are singular points, from the exact expression Eq.(2.28), we found that $M$ may be singular as follows

(1)$[M_1]_2$ and $[M_1]_3$ could have poles in $D_1$ at the zeros of $s_{11}(\lambda)$,

(2)$[M_2]_2$ and $[M_2]_3$ could have poles in $D_2$ at the zeros of $(s^TS^A)_{11}(\lambda)$,

(3)$[M_3]_1$ could have poles in $D_3$ at the zeros of $(S^Ts^A)_{11}(\lambda)$,

(4)$[M_4]_1$ could have poles in $D_4$ at the zeros of $m_{11}(s)(\lambda)$.

We use $\lambda_j(j=1,2\cdots N)$ denote the possible zero point of $M$ in $D_n$, and assume that these possible zeros satisfy the following assumptions.

\begin{assumption}
Assume that

(1)$s_{11}(\lambda)$ has $n_0$ possible simple zeros in $D_1$ denoted by $\lambda_j,j=1,2\cdots n_0$,

(2)$(s^TS^A)_{11}(\lambda)$ has $n_1-n_0$ possible simple zeros in $D_2$ denoted by $\lambda_j,j=n_0+1,n_0+2\cdots n_1$,

(3)$(S^Ts^A)_{11}(\lambda)$ has $n_2-n_1$ possible simple zeros in $D_3$ denoted by $\lambda_j,j=n_1+1,n_1+2\cdots n_2$,

(4)$m_{11}(s)(\lambda)$ has $N-n_2$ possible simple zeros in $D_4$ denoted by $\lambda_j,j=n_2+2,n_2+2\cdots N$,

And these zeros are different, moreover assuming that there is no zero on the boundary of $D_n(n=1,2,3,4)$.
\end{assumption}

\begin{prop}
Let $M_n(n=1,2,3,4)$ be the eigenfunctions defined by (2.14) and assume that the set $\lambda_j(j=1,2\cdots N)$ of singularities are as the above assumption. Then the following residue conditions hold true:
\beq
\mathrm{Re}s_{\lambda=\lambda_j}[M]_2=\frac{m_{33}(s)(\lambda_j)}{\dot{s_{11}(\lambda_j)}s_{21}(\lambda_j)}e^{\theta_{13}(\lambda_j)}[M(\lambda_j)]_1
,\quad 1\leq j\leq n_0;\lambda_j\in D_1.\label{slisp}\eeq
\beq
\mathrm{Re}s_{\lambda=\lambda_j}[M]_3=\frac{m_{32}(s)(\lambda_j)}{\dot{s_{11}(\lambda_j)}s_{21}(\lambda_j)}e^{\theta_{13}(\lambda_j)}[M(\lambda_j)]_1
,\quad 1\leq j\leq n_0;\lambda_j\in D_1.\label{slisp}\eeq
\eqa && \mathrm{Re}s_{\lambda=\lambda_j}[M]_2=\frac{m_{33}(s)(\lambda_j)M_{11}(S)(\lambda_j)-m_{13}(s)(\lambda_j)M_{31}(S)(\lambda_j)}{\dot{(s^TS^A)_{11}(\lambda_j)}s_{21}(\lambda_j)}e^{\theta_{13}(\lambda_j)}[M(\lambda_j)]_1
,\nn\\&&\qquad\qquad\qquad
n_0+1\leq j\leq n_1;\lambda_j\in D_2.\label{slisp}\eeqa
\eqa && \mathrm{Re}s_{\lambda=\lambda_j}[M]_3=\frac{m_{32}(s)(\lambda_j)M_{11}(S)(\lambda_j)-m_{12}(s)(\lambda_j)M_{31}(S)(\lambda_j)}{\dot{(s^TS^A)_{11}(\lambda_j)}s_{21}(\lambda_j)}e^{\theta_{13}(\lambda_j)}[M(\lambda_j)]_1
,\nn\\&&\qquad\qquad\qquad
n_0+1\leq j\leq n_1;\lambda_j\in D_2.\label{slisp}\eeqa
\eqa && \mathrm{Re}s_{\lambda=\lambda_j}[M]_1=\frac{s_{33}(\lambda_j)S_{21}(\lambda_j)-s_{23}(\lambda_j)S_{31}(\lambda_j)}{\dot{(S^Ts^A)_{11}(\lambda_j)}m_{11}(s)(\lambda_j)}e^{\theta_{31}(\lambda_j)}[M(\lambda_j)]_2
\nn\\&&\qquad\quad
+\frac{s_{22}(\lambda_j)S_{31}(\lambda_j)-s_{32}(\lambda_j)S_{21}(\lambda_j)}{\dot{(S^Ts^A)_{11}(\lambda_j)}m_{11}(s)(\lambda_j)}e^{\theta_{31}(\lambda_j)}[M(\lambda_j)]_3
, n_1+1\leq j\leq n_2;\lambda_j\in D_3.\label{slisp}\eeqa
\beq \mathrm{Re}s_{\lambda=\lambda_j}[M]_1=\frac{s_{33}(\lambda_j)[M_(\lambda_j)]_2-s_{32}(\lambda_j)[M_(\lambda_j)]_3}{\dot{m_{11}(s)(\lambda_j)}m_{21}(s)(\lambda_j)}e^{\theta_{31}(\lambda_j)}
, n_2+1\leq j\leq N;\lambda_j\in D_4.\label{slisp}\eeq
where $\dot{f}=\frac{df}{d\lambda}$ and $\theta_{ij}$ given by
\beq \theta_{ij}(x,t,\lambda)=(l_i-l_j)x-(z_i-z_j)t \quad i,j=1,2,3,\label{slisp}\eeq
thus
\beq \theta_{ij}=0,\quad i,j=2,3,\quad\theta_{12}=\theta_{13}=-\theta_{21}=-\theta_{31}=2i\lambda^2x-4ik^2t.\nn\eeq
\end{prop}

Proof: We will only prove (2.39), (2.40) and the other conditions follow by similar arguments. The equation (2.27) means that
\beq M_2=\mu_2e^{(i\lambda^2x+2ik^2t)\hat\sigma} S_2,\label{slisp}\eeq

In view of the expression for $S_2$ given in (2.28), the three columns of Eq.(2.44) read
\beq [M_2]_1=[\mu_2]_1s_{11}+[\mu_2]_2s_{21}e^{\theta_{31}}+[\mu_2]_3s_{31}e^{\theta_{31}},\label{slisp}\eeq
\eqa&&
[M_2]_2=\frac{m_{33}(s)M_{21}(S)-m_{23}(s)M_{31}(S)}{(s^TS^A)_{11}}e^{\theta_{13}}[\mu_2]_1+\frac{m_{33}(s)M_{11}(S)-m_{13}(s)M_{31}(S)}{(s^TS^A)_{11}}[\mu_2]_2
\nn\\&&\qquad\quad
+\frac{m_{23}(s)M_{11}(S)-m_{13}(s)M_{21}(S)}{(s^TS^A)_{11}}[\mu_2]_3,\label{slisp}\eeqa
\eqa&&
[M_2]_3=\frac{m_{32}(s)M_{21}(S)-m_{22}(s)M_{31}(S)}{(s^TS^A)_{11}}e^{\theta_{13}}[\mu_2]_1+\frac{m_{32}(s)M_{11}(S)-m_{12}(s)M_{31}(S)}{(s^TS^A)_{11}}[\mu_2]_2
\nn\\&&\qquad\quad
+\frac{m_{22}(s)M_{11}(S)-m_{12}(s)M_{21}(S)}{(s^TS^A)_{11}}[\mu_2]_3.\label{slisp}\eeqa

Let $\lambda_j\in D_2$ be a simple zero of $(s^TS^A)_{11}(\lambda)$. Solving Eq.(2.45) for $[\mu_2]_2$ and substituting
the result into Eq.(2.46) and Eq.(2.47) yields
\beq [M_2]_2=\frac{m_{33}(s)M_{11}(S)-m_{13}(s)M_{31}(S)}{(s^TS^A)_{11}s_{21}}e^{\theta_{13}}[M_1]_1-\frac{m_{33}(s)}{s_{21}}e^{\theta_{13}}[\mu_2]_1
+\frac{m_{13}(s)}{s_{21}}[\mu_2]_3,\label{slisp}\eeq
\beq [M_2]_3=\frac{m_{32}(s)M_{11}(S)-m_{12}(s)M_{31}(S)}{(s^TS^A)_{11}s_{21}}e^{\theta_{13}}[M_1]_1-\frac{m_{32}(s)}{s_{21}}e^{\theta_{13}}[\mu_2]_1
+\frac{m_{12}(s)}{s_{21}}[\mu_2]_3.\label{slisp}\eeq
Taking the residue of the two equations at $\lambda_j$, we find conditions Eq.(2.39) and Eq.(2.40) in the case when $\lambda_j\in D_2$.

\subsection{ The global relation }

\qquad The spectral functions $S(\lambda)$ and $s(\lambda)$ are not independent which is of important relationship each other. In fact, from Eq.(2.24), we not difficult to find that
\beq \mu_3(x,t,\lambda)=\mu_1(x,t,\lambda)e^{(i\lambda^2 x+2ik^2t)\hat\sigma} S^{-1}(\lambda)s(\lambda),
\lambda\in(D_1\cup D_2,D_3\cup D_4,D_3\cup D_4),\label{slisp}\eeq
as $\mu_1(0,t,\lambda)=\mathbb{I}$, when $(x,t)=(0,T)$, We can evaluate the following relationship which is the global relation
\beq S^{-1}(\lambda)s(\lambda)=e^{-2ik^2T\hat\sigma}c(T,\lambda),\quad
\lambda\in(D_1\cup D_2,D_3\cup D_4,D_3\cup D_4),
\label{slisp}\eeq
where  $c(T,\lambda)=\mu_3(0,t,\lambda)$.

\section{The Riemann-Hilbert problem }

In section 2, we defined the sectionally analytical function $M(x,t,\lambda)$ that its satisfies a Riemann-Hilbert problem which can be formulated in terms of the initial values and boundary values of $\{q(x,t),r(x,t)\}$. For all $(x,t)$, the $\{q_x(x,t),r_x(x,t)\}$ can be recovered by solving this Riemann-Hilbert problem, and the solution $\{q(x,t),r(x,t)\}$ of Eq.(1.4) can be obtained by integration with respect to $x$. So we have the following theorem is established.

\begin{thm}
Suppose that the half-line domain $\Omega=\{0<x<\infty,0<t<T\}$ with sufficient smoothness and decays as $x\rightarrow\infty$, and assume that $\{q(x,t),r(x,t)\}$ is a solution of Eq.(1.4) in half-line domain $\Omega$ which can be reconstructed from the initial value $\{q_0(x),r_0(x)\}$ and boundary values $\{g_0(t),h_0(t),g_1(t),h_1(t)\}$ lie in the Schwartz class defined as follows.
\eqa\begin{array}{l}
q_0(x)=u(x,t=0),\; r_0(x)=v(x,t=0),\\
g_0(t)=q(x=0,t),\; h_0(t)=r(x=0,t),\\
g_1(t)=q_x(x=0,t),\; h_1(t)=r_x(x=0,t),
\end{array}\label{slisp}\eeqa
like Eq.(2.24) using the initial data and boundary data to define the spectral functions $s(\lambda)$ and $S(\lambda)$,further defining the jump matrix $J_{m,n}(x,t,\lambda)$. Assume that the zero point of the $s_{11}(\lambda),(s^TS^A)_{11}(\lambda),(S^Ts^A)_{11}(\lambda)$ and $m_{11}(s)(\lambda)$ is $\lambda_j(j=1,2\cdots N)$ are as in assumption 2.5, that is the following assumptions.

Then the $\{q_x(x,t),r_x(x,t)\}$ of Eq.(1.4) is
\eqa\begin{array}{l}
q_x(x,t)=-2i\lim_{\lambda\rightarrow\infty}(\lambda M(x,t,\lambda))_{12},\\
r_x(x,t)=-2i\lim_{\lambda\rightarrow\infty}(\lambda M(x,t,\lambda))_{13},
\end{array}\label{slisp}\eeqa
where $M(x,t,\lambda)$ satisfies the following $3\times 3$ matrix Riemann-Hilbert problem:

(1)$M$ is a sectionally meromorphic on the Riemann $\lambda$-sphere with jumps across the contours on $\bar D_n\cap\bar D_m(n,m=1,2,3,4)$ (see figure 2).

(2)$M$ satisfies the jump condition with jumps across the contours on $\bar D_n\cap\bar D_m(n,m=1,2,3,4)$
\beq M_n(\lambda)=M_mJ_{m,n},\quad \lambda\in \bar D_n\cap \bar D_m,n,m=1,2,3,4;n\neq m.\label{slisp}\eeq

(3)$M(x,t,\lambda)=\mathbb{I}+O(\frac{1}{\lambda}),\quad \lambda\rightarrow\infty.$

(4)The residue condition of $M$ is showed in Proposition 2.6.

\end{thm}
Proof: We can use similar method with\cite{Xu2013} to prove this Theorem. It only need to prove Eq.(3.2) and this equation follows from the large $\lambda$ asymptotics of the eigenfunctions.

Thus, the solution of the coupled Fokas-Lenells equations $\{q(x,t),r(x,t)\}$ can be obtained by integration with respect to $x$.

\section{Conclusions and discussions}

In this paper, we consider IBV of the CFL equation on the half-line. Using the unified transform method for nonlinear evolution systems which taking the form of Lax pair isospectral deformations and whose corresponding continuous spectra Lax operators, assume that the solutions $q(x,t)$ and $r(x,t)$ exists, we show that it can be represented in terms of the solution of a matrix Riemann-Hilbert problem formulated in the plane of the complex spectral parameter $\lambda$. For other $3\times3$ matrix Lax pair integrable equations, can we construct their solution of a matrix Riemann-Hilbert problem formulated in the plane of the complex spectral parameter $\lambda$ by the similar method?
In paper \cite{XU2015}, Xu and Fan use the Deift-Zhou method to studied the long-time asymptotics for the solutions of the decay initial value on the full-line. Moreover, under the assumption that the initial and boundary values lie in the Schwartz class, Chen and Yan have successfully applied the nonlinear steepest descent method to analyze the long-time asymptotic for the solution of decay IBV problem of the FL equation on the half line in \cite{Chen2017}, can we do the long-time asymptotics for the solutions of the decay initial and boundary values of CFL equations following the same ways as for the DP equation \cite{monvel2015}? These questions will be discussed in our future work.

\subsection*{Acknowledgements}

This work is partially supported by the National Natural Science Foundation of China under Grant Nos. 12271008  and  11601055,
Natural Science Foundation of Anhui Province under Grant No.1408085QA06.

\end{document}